\documentclass{aa}  

\usepackage{graphicx}
\usepackage{txfonts}
\begin{document} 

   \title{Ammonia-methane ratios from H-band near-infrared spectra of late-T and Y dwarfs}


   \author{
       E.\ L.\ Mart\'in \inst{1,2,3}
                  \and  J.-Y.\ Zhang \inst{1} 
               \and
      P.\ Esparza \inst{4}
        \and     F.\ Gracia \inst{1}
       \and     J.\ L.\ Rasilla \inst{1}       
            \and
                   T. Masseron \inst{1,2}  
                      \and  A.\ J.\ Burgasser \inst{5} 
        }

   \institute{Instituto de Astrof\'isica de Canarias (IAC), Calle V\'ia L\'actea s/n, E-38200 La Laguna, Tenerife, Spain \\
       \email{ege@iac.es}
       \and
       Departamento de Astrof\'isica, Universidad de La Laguna (ULL), E-38206 La Laguna, Tenerife, Spain
       \and
       Consejo Superior de Investigaciones Cient\'ificas (CSIC), E-28006 Madrid, Spain
      \and
      Departamento de Qu\'imica, Universidad de La Laguna (ULL), E-38206 La Laguna, Tenerife, Spain
            \and
      Center for Astrophysics and Space Science, University of California San Diego, 9500 Gilman Drive, La Jolla, CA 92092, USA
       }

   \date{\today{},\today{}}

 
  \abstract
   {
   }
   {Our goals are to investigate the relative absorption strengths of ammonia and methane using low-resolution H-band (1.5-1.7 microns) spectra obtained in the laboratory and compared with observational spectra of late-T and Y dwarfs, and to estimate what can be expected from the wide-angle low-resolution near-infrared spectroscopic survey that will be provided by the upcoming Euclid space mission.}
   {Gas cells containing ammonia and methane at atmospheric pressure were custom-made in our chemical laboratory. Low-resolution near-infrared spectroscopy of these gas cells were collected in our optical laboratory. They are compared with simulated spectra using the high-resolution transmission molecular absorption database (HITRAN) for temperatures of 300 K and 500 K, and with near-infrared spectra of late-T and Y dwarfs, Jupiter and Saturn. We selected for this investigation the spectral region between 1.5 and 1.7 microns (H-band) because it is covered by the Euclid red grism, it is particularly sensitive to the relative proportions of ammonia and methane opacity, and it is free from strong contributions of other abundant molecules such as water vapour. 
   }
   {The laboratory spectra showed that the ammonia and methane features that are present in the simulations using the HITRAN database are incomplete. Using our laboratory spectra, we propose a modified version of the NH3-H spectral ratio with expanded integration limits that increases the amplitude of variation of the index with respect to spectral type.  
   Combinations of our laboratory spectra were used to find the best fits to the observed spectra with the relative absorption ratio of ammonia to methane as a free parameter. A relationship was found between the T$\rm_{eff}$ and the ratio of ammonia to methane from spectral classes T5 to Y2 (1100 K -- 350 K), in qualitative fairly good agreement with theoretical predictions for high gravity objects and temperatures from 1100 K to 500 K. The ammonia to methane ratios in late-T and Y dwarfs are similar to that of Jupiter, suggesting a similar chemical composition. Simulations of the spectroscopic performance of Euclid suggest that it will yield T$\rm_{eff}$ values and ratios of ammonia to methane for over 10$^3$ late T dwarfs in the whole wide survey.}
  {} 

\titlerunning{Ammonia-methane ratios in late-T and Y dwarfs}
   \maketitle
%
%
%
\section{Introduction}
\label{esdT:intro}

Brown dwarfs (BDs) do not stabilize on the Hydrogen (H) burning main-sequence, and hence they keep cooling down to very low temperatures. The coolest BDs have been spectroscopically classified as T dwarfs and Y dwarfs, based on the presence of methane and ammonia in near-infrared spectra, respectively \citep{burgasser02,delorme08a,cushing11}, although the presence of ammonia has been established not only in Y dwarfs but also in late-T dwarfs \citep{bochanski11, canty15}.  

The Wide Field Infrared Survey Explorer
\citep[WISE;][]{wright10} has been very effective to discover Y dwarfs and low-resolution spectroscopic follow-up has been carried out from the ground and from space. In particular, homogeneous sets of low-resolution near-infrared spectra have been collected with Keck Near Infrared Spectrometer (NIRSPEC) and with the Hubble Space Telescope (HST) for late-T and Y dwarfs down to spectral class Y2 
\citep{mclean03,schneider2015, cushing2021}. 

Estimates of the effective temperature (T$\rm_{eff}$) of Y dwarfs are difficult because theoretical fits to their spectral energy distribution are of modest quality, which indicates that the chemistry calculations and/or the sources of opacity may be incomplete, and values estimated from their luminosity and assumed radii using the Stefan-Boltzmann law are hampered by uncertainties due to possible effects of unresolved binarity \citep{leggett13, luhman16a}. 

The appearance of ammonia in the near-infrared spectra of ultracool dwarfs has been discussed by several authors \citep{saumon00, leggett07a, warren07c, bochanski11, burgasser12a, canty15}.  
A spectral index named NH3-H was defined by \citet{delorme08a} to quantify the effects of ammonia absorption from 1.53 to 1.56 microns. Their NH3-H index has been adopted for spectral type quantification from T5 to Y2 dwarfs \citep{cushing2021}. 

The upcoming Euclid\footnote{http://sci.esa.int/euclid/} space mission is expected to provide the next order of magnitude leap in the numbers of detected ultracool dwarfs. At the end of six years of nominal mission, the Euclid wide survey is expected to reach 15,000 square degrees of the extragalactic sky observed at a single epoch, and the Euclid deep surveys are expected to cover between 40 and 50 square degrees at multiple epochs. The Near Infrared Spectrometer and Photometer (NISP) is a two-channel instrument for the Euclid space mission. The photometric channel is equipped with three broad-band filters, and the spectroscopic channel is equipped with three identical low-resolution grisms covering the spectral range from 1250 nm to 1850 nm with a spectral resolving power of R = 380, and one grism covering the spectral range from 920 nm to 1250 nm. The latter grism is expected to be used only for the Euclid Deep Fields, and not for the wide survey. The spectroscopic observations will be made in survey mode without any slit.  The NISP field of view is 0.53 square degrees covered with a matrix of 4x4 Teledyne detectors, each of them with 2040x2040 18 micron pixels with a scale of 0.3 arcsec per pixel.    

Updated simulations of the number counts of ultracool dwarfs in the Euclid wide survey using the most recent information on the photometric passbands were provided in \citet{solano2021}. In this paper we will provide updated simulations for the number of T and Y dwarfs expected to be detected with the NISP spectroscopic mode in the Euclid wide survey. 
We will show that Euclid will provide useful low-resolution H-band spectra for an unprecedented large number of T dwarfs ($\ge10^3$ for T5-T8 subclasses). Consequently, it is important to investigate efficient methods to derive T$\rm_{eff}$ values for T dwarfs using NISP spectra. Our first goal is to confirm that the NH3-H spectral index is really an indicator of the absorption of ammonia in late-T and Y dwarfs, and to adjust its integration limits using our laboratory spectra. Our second goal is the determination of the ammonia to methane ratio as a probe to the atmospheres of ultracool dwarfs using low-resolution H-band spectra.

This paper presents laboratory spectra of gas cells filled with ammonia and methane, which were made with the aim of helping with the identification of these molecules in the atmospheres of ultracool dwarfs.  The rest of the paper is organized as follows: 
Section \ref{Y:Lab_spec} describes the experimental setup for the laboratory measurements.
Section \ref{Y:Comparison1} compares the laboratory spectra with simulated spectra using the HITRAN database and redefines the NH3-H index. 
Section \ref{Y:Comparison2} compares the laboratory spectra with observed near-infrared  spectra from IRTF, Hubble, and Keck for dwarfs with spectral types from T5 to Y2, and for Jupiter and Saturn. Ammonia to methane ratios are derived using a best matching algorithm between the laboratory and the observed spectra.  
Section \ref{Y:discussion} summarizes the results and the implications. 
 
%
 
 %
%
%
\section{Low-resolution laboratory spectra}
\label{Y:Lab_spec}

Custom made gas cells were made to fit into the spectrophotometer Cary 5000 located at the optical laboratory of the Instituto de Astrofísica de Canarias (IAC). They are G2 infrasil cells from International Crystal Lab, 13 cm long and with window size 38 mm in diameter and 6 mm thick. 
 
The cells were filled with different molecular gas using a vacuum pump system in our  laboratory located at the chemistry department of the Universidad de La Laguna. The procedure for filling the cells with gas is described in more detail in 
\citet{valdivielso10}. 
The gas cells with ammonia (NH3), argon (Ar), and methane (CH4) were completely filled with each of those gases at the atmospheric pressure. 

The gas cells were taken to the IAC and they were left to settle to the ambient conditions inside the Cary 5000 for at least five hours before taking the measurements. The Cary 5000 has two channels that allow for two gas cells to be measured simultaneously. In one channel we placed a gas cell filled with ammonia or methane, and in the other channel we placed an identical gas cell filled with Ar so that the instrumental response would be corrected by the software of the instrument. 
 
The laboratory spectra of all the gas cells were taken on 8/8/2021 with an integration time of 0.2 s for each step of 0.5 nm. The total time to measure a spectrum from 800 nm to 3000 nm was 6.3 hours. The ambient conditions in the laboratory were temperature 298 K and humidity 57 \% .  
%
\begin{figure*}
   \centering
\includegraphics[width=\linewidth]{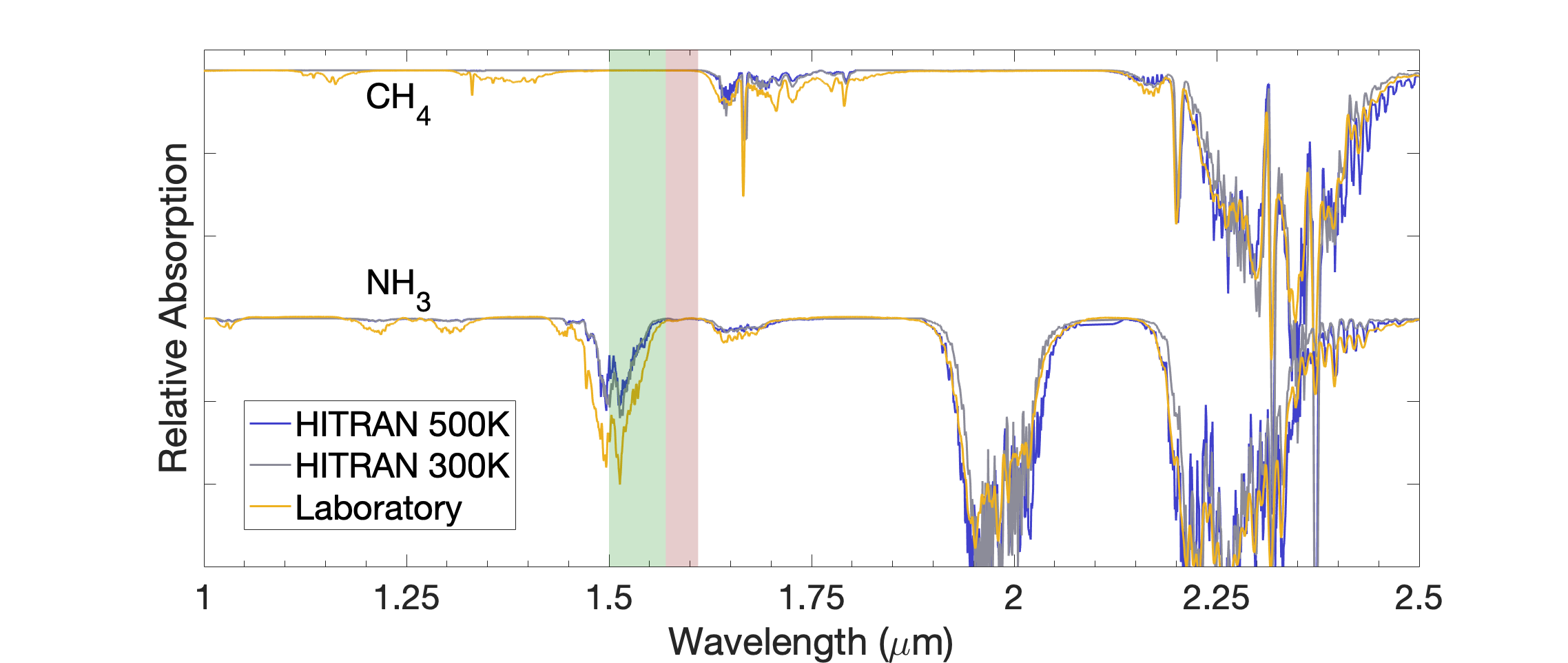}
   \caption{
Our laboratory spectrum for ammonia and methane compared to simulated spectra of the same molecules using HITRAN for temperatures of 500 K and 300 K. The wavelength regions proposed for the new NH3-H index are marked in color bands: green for the numerator, red for the denominator.  }
        \label{fig_Y:hitran}
\end{figure*}

\section{Comparison between laboratory and simulated spectra}
\label{Y:Comparison1}
Our laboratory spectrum for ammonia and methane were compared with simulated spectra for the same molecules  using the HITRAN database \citep{gordon2017hitran2016} 
for temperatures of 300 K and 500 K. The comparisons in the full spectral range from 1.0 to 2.5 microns are shown in Fig.\ \ref{fig_Y:hitran}. The agreement is quite good between 2.0 and 2.5 microns, but the differences increase towards shorter wavelengths.  This is not a surprise because it has been reported before that HITRAN is very incomplete for hot bands and high-J molecular transitions \citep{bailey2012modelling,wong2019atlas}.
 
We use our laboratory spectrum of ammonia to propose a redefinition of the integration limits for the spectral index NH3-H, which was originally defined using HITRAN opacity data \citep{delorme08a}. These authors used as numerator the spectral region from 1.53 to 1.56 microns, and as denominator the region from 1.57 to 1.60 microns. Based on the appearance of the ammonia band in our laboratory spectra, we propose to redefine the NH3-H index using as numerator the region from 1.50 to 1.57 microns, and as denominator the region from 1.57 to 1.61 microns. These integration limits are shown in the bottom part of Fig.\ \ref{fig_Y:hitran}. They sample a region dominated by the ammonia absorption and not by methane. 

\begin{figure}
   \includegraphics[width=\linewidth, angle=0]{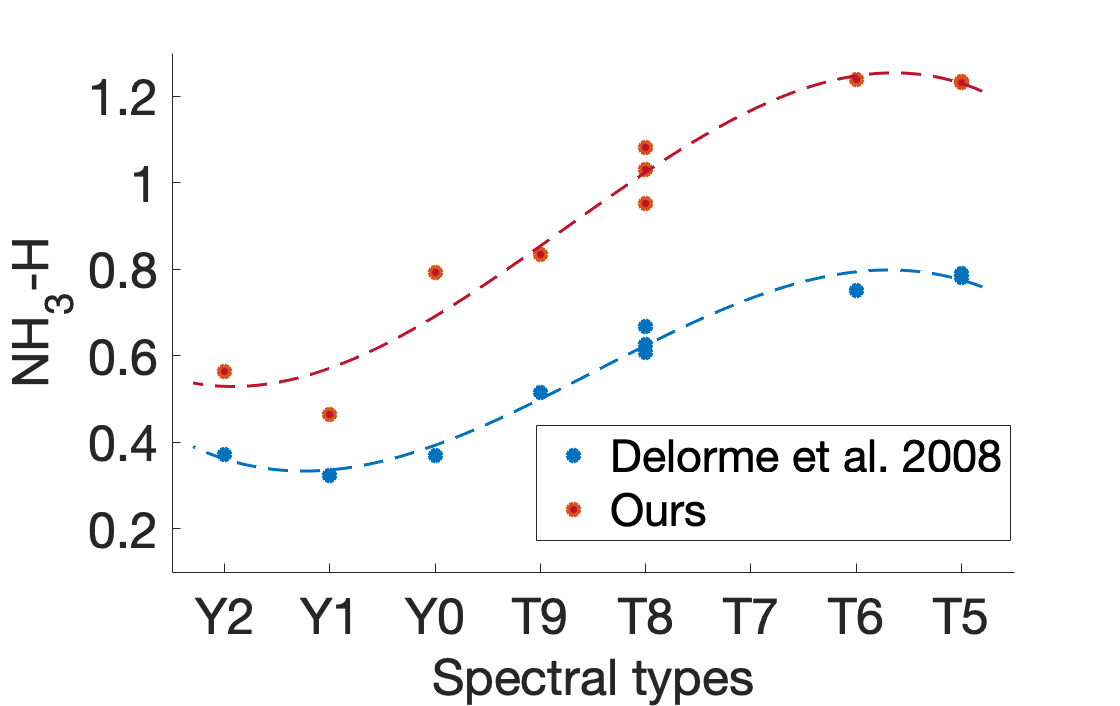}
   \caption{The redefined NH3-H index using our laboratory spectra has a higher sensitivity to spectral class (larger amplitude of variation) than the original definition by \citet{delorme08a} based on HITRAN data. Third order polynomial fits to the relationship between the indeces and the spectral types are denoted with dashed lines of different colors as labelled.  }
     \label{fig_Y:NH3_H_index_compare}
\end{figure}

The increased wavelength range chosen for our proposed modification of the NH3-H spectral index is motivated by a better understanding of the ammonia absorption in this region and it has the advantage of increasing the flux collected in each passband for the classification of ultra faint objects of T and Y type detectable with NISP. 
The comparison between previous NH3-H index and ours of dwarfs from spectral type T5 to Y2 are shown in Fig.\ \ref{fig_Y:NH3_H_index_compare}.

\section{Comparison between laboratory and HST spectra}
\label{Y:Comparison2}
\begin{figure}
   \includegraphics[width=\linewidth, angle=0]{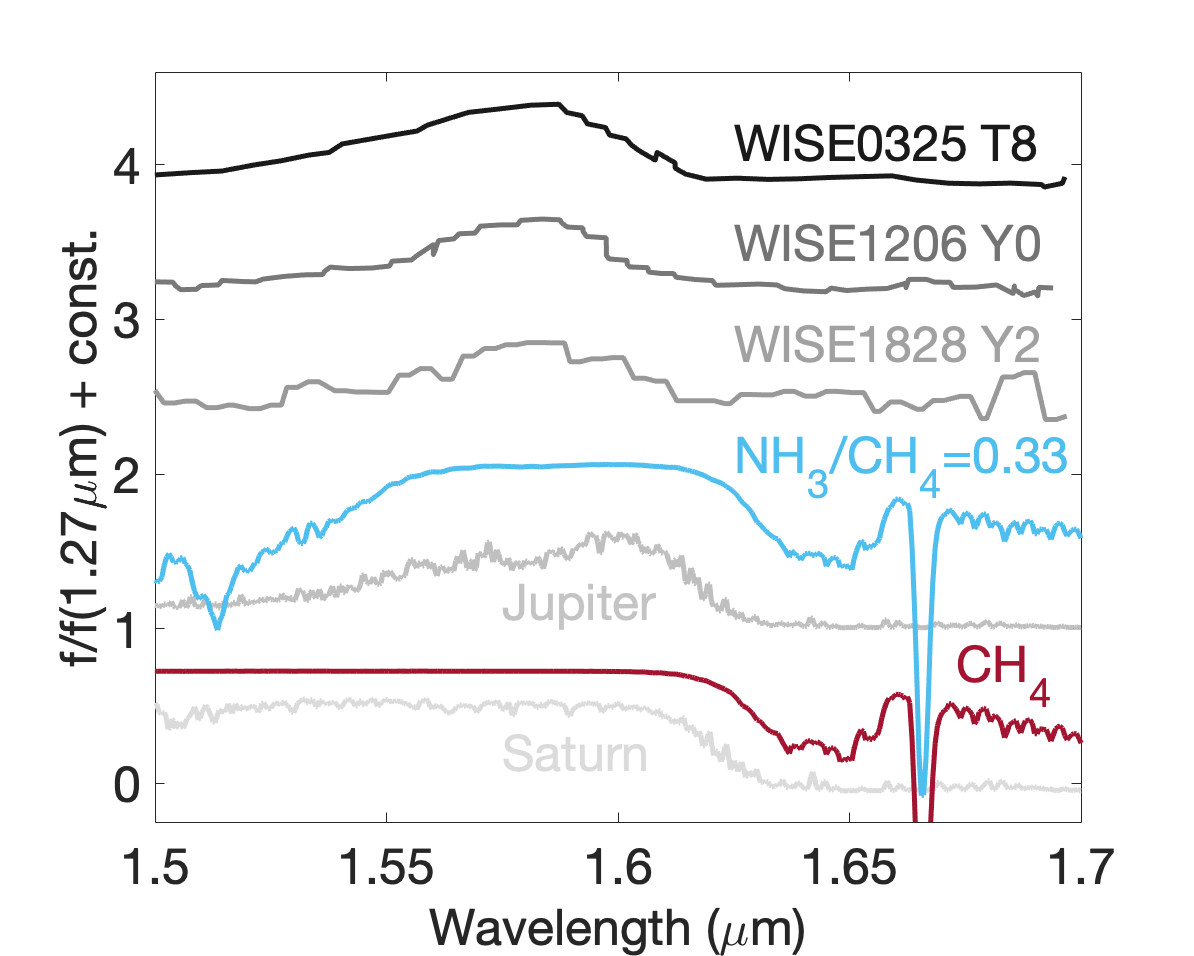}
   \caption{
Our laboratory spectrum for the typical best fitting combination of ammonia and methane, and pure methane, compared to HST spectra of a T8, two Y dwarfs and two giant gas planets. 
   }
     \label{fig_Y:plot_HST}
\end{figure}

Different combinations of the relative strengths of ammonia and methane were computed using our laboratory spectra. An example of the typical combination that provides the best fit is shown in Fig.\ \ref{fig_Y:plot_HST} together with HST spectra of a T8 dwarf and two Y dwarfs \citep{schneider16a, cushing14}. This combination, denoted with a blue line, corresponds to the case where we scaled our laboratory spectrum of ammonia by a factor of 0.33 and we added it to our laboratory spectrum of methane.

The spectral region from 1.5 to 1.7 microns is particularly sensitive to the effects of changing the relative contribution of ammonia with respect to methane. Furthermore this spectral region is thought to be free from significant contribution of other molecules such as water vapour \citep{delorme08a, canty15}. 

\begin{figure}
   \includegraphics[width=\linewidth, angle=0]{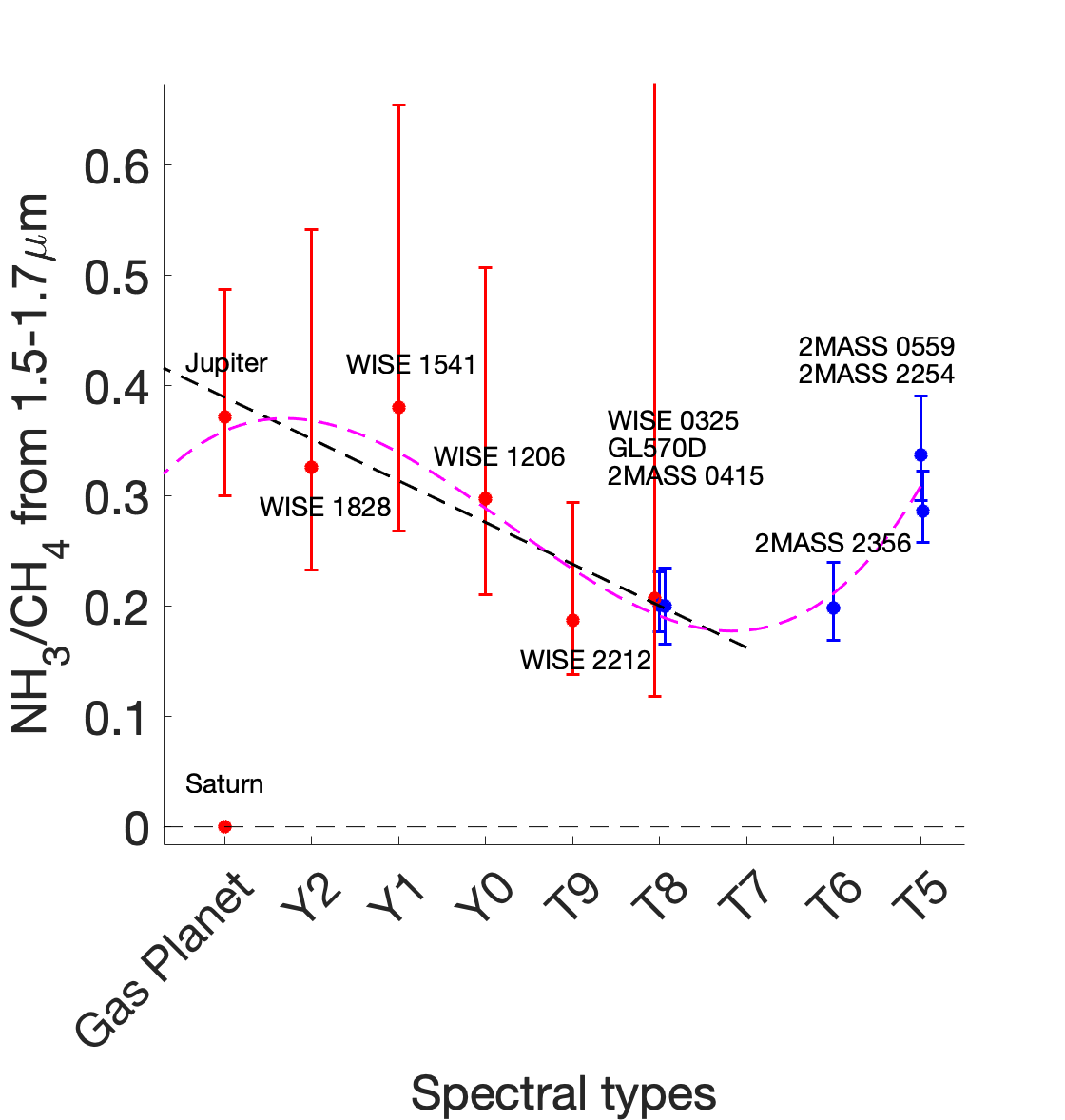}
   \includegraphics[width=\linewidth, angle=0]{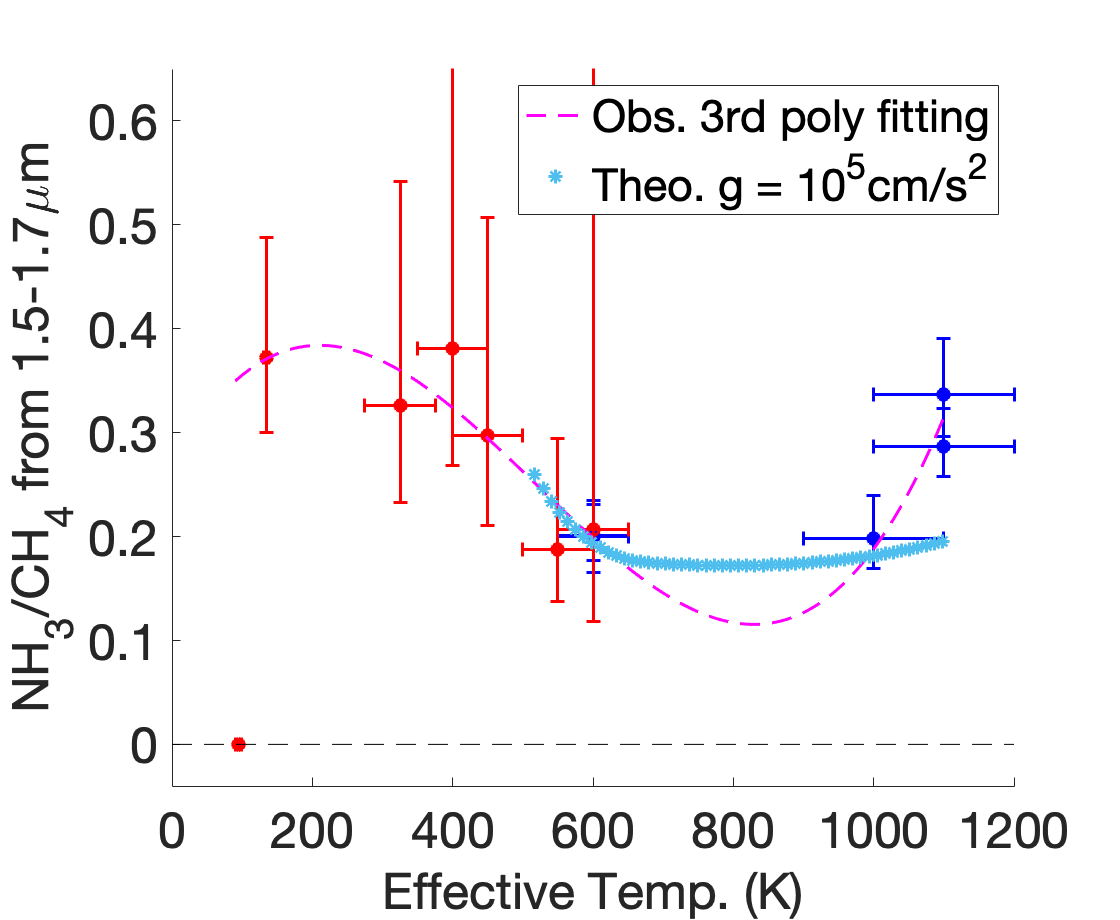}
   \caption{
Ammonia to methane ratios derived for late T dwarfs and Y dwarfs with 95\% confidence level, from NIRSPEC data (blue) and HST data (red) compared with Jupiter and Saturn from IRTF data (red) as functions of spectral type (upper panel) and T$_{\rm eff}$ (lower panel).  The T$_{\rm eff}$ of T, Y dwarfs, Jupiter and Saturn adopted in this work are listed in Table 1. The pink dashed curves are the 3rd-degree polynomial fittings for all object (except Saturn) against spectral type and T$_{\rm eff}$ and the black dashed line is the linear fitting later than T8. The results are compared with the scaled theoretical ratios estimated from \citet{2014ApJ...797...41Z}, which are shown in light blue color.
   }
     \label{fig_Y:plot_ratio}
\end{figure}

Least square fit and a confidence interval estimate were computed using the \verb|lsqcurvefit| and \verb|nlparci| routines available in the Optimization Toolbox and Statistics and Machine Learning Toolbox in the MATrix LABoratory (MATLAB). We found that the best fit to the HST and NIRSPEC spectra of late-T and Y dwarfs in the spectral window from 1.5 to 1.7 microns were achieved with the ratios of ammonia to methane shown in Fig.\ \ref{fig_Y:plot_ratio}. 

Using the same method as described above, we also derived the ammonia to methane ratio in the spectra of Jupiter and Saturn that we downloaded from the NASA Infrared Telescope Facility (IRTF) archive \citep{vacca03}. Our ratio for Jupiter is plotted in Fig.\ \ref{fig_Y:plot_ratio} and it is consistent (within the 1 sigma uncertainty) with that found in the literature using models and observations of the Galileo spacecraft (0.26, \citet{taylor07}). 
Our ratio for Saturn agrees with \citet{owen1977study} who reported on non-detection of ammonia in high-resolution spectra of Saturn obtained at Palomar observatory in the wavelength region relevant for this work.  

In Fig.\ \ref{fig_Y:plot_ratio} (lower panel) we show the comparison of our ammonia to methane ratios with the theoretical predictions for high-gravity objects ($g=10^5 \mathrm{cm\cdot s^{-2}}$) from \citep{2014ApJ...797...41Z}. To compare the behaviour of the models with other data, we used the ratios of theoretical column densities scaled by a constant factor to make two values reaching an agreement at 600K.  There is a good qualitative agreement in the sense that the models predict a ratio that has a minimum around 800 K and increases towards cooler and hotter temperatures.

%

%
\section{Discussion and final remarks}
\label{Y:discussion}

Our laboratory spectra of ammonia and methane contain absorption features that are too weak or not present in the HITRAN database. We used our laboratory spectrum of ammonia to redefine the spectral index NH3-H in late-T and Y  dwarfs. This index is a sensitive indicator of the spectral type of ultracool brown dwarfs and it can be used to calibrate the data provided by the slitless spectroscopic mode of the NISP instrument on Euclid. 
Our new defined NH3-H index has broader integration limits and an amplitude of variation with spectral type almost twice larger than the one based on HITRAN and originally proposed by \citet{delorme08a} from spectral type T5 to Y2.

There is a smooth relation between the ammonia to methane ratio derived from our best fits to the observed spectral region from 1.5 to 1.7 microns and T$_{\rm eff}$ of the objects considered, which is qualitatively in agreement with theoretical predictions taken from the ratio of the theoretical column densities of ammonia and methane under a high surface gravity environment ($g=10^5 \mathrm{cm\cdot s^{-2}}$) \citep{2014ApJ...797...41Z}. All the late T and Y dwarfs considered in this work have ammonia to methane ratios consistent with that of Jupiter within the uncertainties, suggesting that they have similar chemical abundances. 

\begin{figure}
   \includegraphics[width=\linewidth, angle=0]{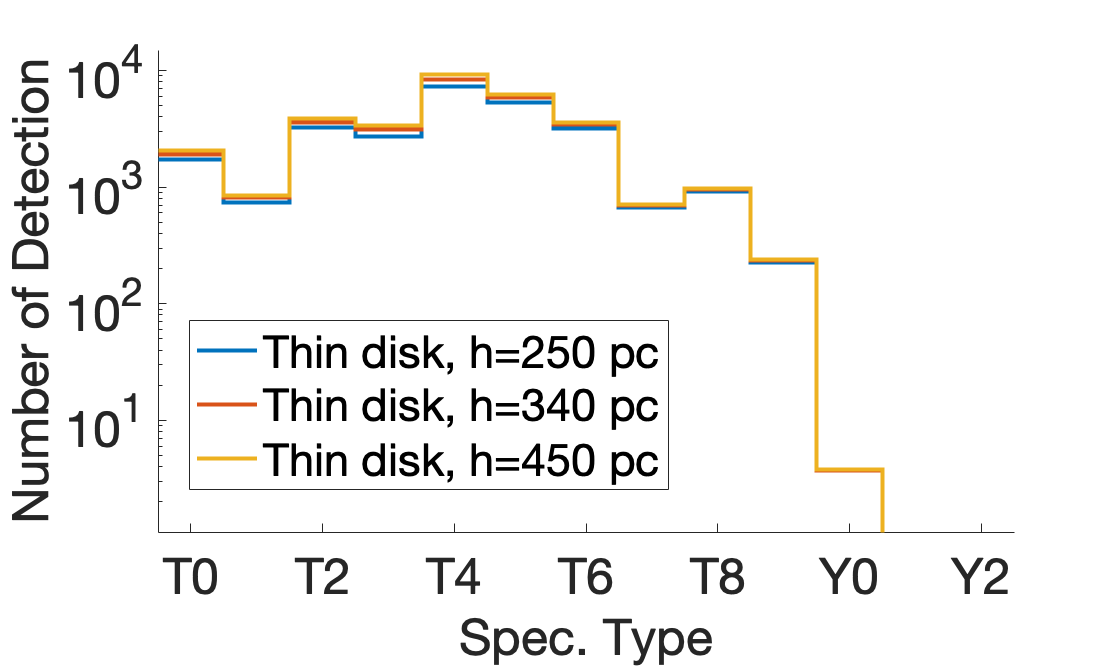}
   \caption{
Simulated number counts of T and Y dwarfs detectable (SNR$\ge$10) in the full Euclid wide survey with the spectroscopic mode of NISP for different values of the galactic scale height of the brown dwarf population in the think disk of the Galaxy.  }
     \label{fig_Y:plot_Euclid}
\end{figure}

Using the same assumptions as in \citet{solano2021} and the densities of objects of different spectral types in the solar neighborhood \citet{kirkpatrick21b}, we provide here estimates of the numbers of dwarfs with spectral types from T0 to Y2 that could be detected by NISP spectroscopy with SNR$>$10 in the Euclid wide survey (Fig.\ \ref{fig_Y:plot_Euclid}). Note that the numbers of T5 to T8 dwarfs spectroscopically characterized by NISP are expected to be $\ge$10$^3$, and they rapidly decrease into the Y dwarf domain due to the intrinsic faintness at near-infrared wavelengths of those sources. The results of this work indicate that ammonia to methane ratios can be measured in late-T and Y dwarfs using low-resolution near-infrared spectra such as those that will be provided by Euclid/NISP. These ratios measured in a large sample with NISP will be useful to explore the diversity of chemical and physical properties that are likely to exist among the coolest objects in the solar vicinity.

\begin{table}
\centering
 \caption[]{
Adopted T$_{\rm eff}$ values and uncertainties for the different spectral types and giant gas planets considered in this work.
}
 \begin{tabular}{@{\hspace{0mm}}c 
 @{\hspace{2mm}}c @{\hspace{2mm}}c@{\hspace{0mm}}}
 \hline
 \hline
Type       &  T$_{\rm eff}$ & References  \cr
 \hline
            &    (K)           &            \cr
 \hline
T5  &  1100$\pm$100    & \citet{dupuy17}   \cr
T6  &  1000$\pm$100    & \citet{dupuy17}  \cr
T8  &  600$\pm$50    & \citet{schneider2015}  \cr
T9  &  550$\pm$50    & \citet{schneider2015} \cr
Y0  &  450$\pm$50    & \citet{schneider2015}  \cr
Y1  &  400$\pm$50    & \citet{schneider2015}   \cr  
Y2  &  300$\pm$50    &  \citet{cushing2021} \cr 
Jupiter  &  134$\pm$4    &  \citet{aumann1969internal} \cr 
Saturn  &  97$\pm$4    &  \citet{aumann1969internal} \cr
 \hline
 \label{tab_Y:sample}
 \end{tabular}
\end{table}

\begin{acknowledgements}
ELM acknowledges  support  from  the Agencia  Estatal  de  Investigaci\'on  del  Ministerio  de  Ciencia  e Innovaci\'on (AEI-MCINN) under grant PID2019-109522GB-C53\@.
J.-Y. Zhang acknowledges a summer grant from the Instituto de Astrofisica de Canarias. 
ELM thanks Mike Cushing for sending the HST infrared spectrum published
in \citet{cushing2021}. 
This research has made use of the Simbad database, operated
at the centre de Donn\'ees Astronomiques de Strasbourg (CDS), and
of NASA's Astrophysics Data System Bibliographic Services (ADS).
 
\end{acknowledgements}
%

%
%

%
%
\bibliographystyle{aa} 
\bibliography{mnemonic,abiblio} 

%
%

%



\end{document}